\documentclass[aps,prb,twocolumn,showpacs,preprintnumbers]{revtex4}
\usepackage{graphicx}
\usepackage{bm}
\usepackage{color}
\usepackage{epsfig}
\usepackage{epstopdf}

\begin{document}
\title{Spin-orbit coupling, strong correlation, and insulator-metal transitions:\\ 
 the J$_{\rm eff} = \frac{3}{2}$ ferromagnetic Mott insulator Ba$_{2}$NaOsO$_{6}$}
\author{Shruba Gangopadhyay and Warren E. Pickett }
\affiliation{Department of Physics, University of California, Davis CA 95616, USA}

\begin{abstract}
The double perovskite Ba$_{2}$NaOsO$_{6}$ (BNOO), an exotic example of a very high
oxidation state
(heptavalent) osmium $d^1$ compound and also uncommon by being a ferromagnetic Mott
insulator without Jahn-Teller (JT) distortion,
is modeled using the density functional theory (DFT) hybrid functional
based exact exchange for correlated
electrons (oeeHyb) method 
and including spin-orbit coupling (SOC). 
The experimentally observed narrow gap ferromagnetic
insulating ground state is obtained, with easy axis along [110] in accord with
experiment, providing support that this approach provides a realistic method
for studying this system. The predicted spin density for [110] spin orientation
is nearly cubic (unlike for other directions), providing an explanation for the
absence of JT distortion. An orbital moment of -0.4$\mu_B$ 
strongly compensates the +0.5$\mu_B$ spin moment on Os, leaving a strongly compensated moment
more in line with experiment. Remarkably, the net moment lies primarily on 
the oxygen ions. 
An insulator-metal
transition by rotating the magnetization direction with an external field under moderate
pressure is predicted as one consequence of strong SOC, and metallization under moderate pressure 
is predicted.
Comparison is made with the isostructural, isovalent
insulator Ba$_2$LiOsO$_6$ which however orders antiferromagnetically. 
\end{abstract}
\date{\today}
\maketitle

\section{Introduction}
Over the last three decades the orbital physics of  $\mathit{d}^{1}$ systems and its interplay with 
spin has focused primarily on transition metal oxides, especially 3$\mathit{d}$ transition metals (TMs). 
While the magnetic $\mathit{d}^{1}$ configuration has been studied mostly in $3d$ systems, it can also occur in 
mid- to late-5$\mathit{d}$ TM ionic systems. For these heavy ions spin-orbit coupling (SOC) becomes a competing 
factor, mixing the various spin, orbital, charge, and lattice degrees of freedom. The interplay of 
strong electron correlation and large SOC is relatively less explored, and certainly not well understood 
at all, because the behavior involves so many comparable energy scales. This situation arises in a broad 
family of magnetic Mott insulating systems in which three-fold degenerate $t_{2g}$ orbitals 
are partially filled.\cite{balentsprb1} In such systems orbital degeneracy is 
protected only by cubic lattice symmetry, 
and typically the crystal field splitting is large enough that $e_{g}$ orbitals are out of the picture. 

In $5d$ $t_{2g}$ subshells where SOC remains unquenched (cubic symmetry), 
the six one-electron levels split into an upper 
$J$=1/2 doublet and a lower $J$=3/2 quartet. In this category Ir-based based magnets 
have been studied actively.\cite{ir1, ir2, ir3, ir4, ir5} A prominent class of such 
systems is the ordered double perovskites, 
with chemical formula A$_{2}$BB$'$O$_{6}$. We are interested in the case where B is a closed shell cation
and B$'$ is a magnetic ion; in such cases unusually high formal valence states can arise. A few examples 
attracting recent interest are
B$'$=Ru$^{5+}$ and Os$^{5+}$ in A$_{2}$NaB$'$O$_{6}$ (A=La and Nd),\cite{Nd2NaOsO6,la2nabo6prl, la2NaOs_ruO6} 
Mo$^{5+}$ in Ba$_{2}$YMoO$_{6}$, 
Os$^{6+}$ in Ba$_{2}$CaOsO$_{6},$\cite{bcco} 
and heptavalent Os in Ba$_{2}$BOsO$_{6}$ (B =Li, Na). 
If we narrow our focus to $\mathit{d}^{1}$ B$'$ ions only, 
the possibilities are practically confined to Mo$^{5+}$, Re$^{6+}$ and 
Os$^{7+}$).  Song {\it et al.} have reported a theoretical study\cite{song2014} of KOsO$_4$
with heptavalent Os, where large SOC,
strong correlations, and structural symmetry breaking conspire to produce an unexpectedly large
orbital moment in the $e_g^1$ shell that nominally supports no orbital moment.

The recently studied compounds\cite{bnoobloo,fisherbnoo} Ba$_{2}$BOsO$_{6}$ (B=Li, Na) 
show many features to make them of current interest. Besides the double-perovskite structure, 
and being a rare example of a heptavalent osmium compound, Ba$_2$NaOsO$_6$ (BNOO) is exotic in
being a {\it ferromagnetic Mott insulator},\cite{crystalbnoo,fisherbnoo} 
with order appearing at T$_C$ =6.8K with Curie-Weiss temperature $\Theta_{CW}=$ -10K. 
Although its single $t_{2g}$ electron orders magnetically, it shows no evidence of the anticipated
orbital order that causes Jahn-Teller distortion and should destroys its cubic symmetry. The sister compound 
La$_{2}$NaOsO$_{6}$, on the other hand, with high-spin  $\mathit{d}^{3}$ Os configuration and a 
nominally cubic symmetry, is observed to be highly distorted.\cite{os_arrangment} This distortion 
is ascribed to geometrical misfit arising from incompatible ionic radii. 

There is a recent example of 
an Os-based based $5d^4$ perovskite compound BaOsO$_{3}$ that remains cubic;\cite{baoso3} 
on the other hand a related 
perovskite $5d^5$ NaOsO$_{3}$ does distort.\cite{naoso3} The question of origin of the magnetic ordering in BNOO is
surely a delicate one, since isostructural, isovalent, and also Mott insulating
Ba$_{2}$LiOsO$_{6}$ (BLOO) orders 
{\it antiferromagnetically} in spite of a very similar Curie-Weiss susceptibility\cite{crystalbnoo} 
and similar volume. 

Lee and Pickett demonstrated\cite{Pickett_bnoo} that, 
before considering magnetism and on-site interaction effects, SOC
splits the $t_{2g}$ bands into a lower $J$=$\frac{3}{2}$ quartet and an upper $J$=$\frac{1}{2}$ doublet,
as expected. Since BNOO is observed to be insulating and effects of spin-orbit coupling
drive the behavior, it provides the first
``J$_{\rm eff}$=$\frac{3}{2}$'' Mott insulator at quarter-filling, 
analogous to the ``J$_{\rm eff}$=$\frac{1}{2}$''
Mott insulators at half-filling that are being studied in $5d^5$ systems.\cite{ir4,jeff2,jeff3}

Including spin polarization and on-site Hubbard U repulsion beyond the semilocal density
approximation (DFT+U+SOC) with both the Wien2k and FPLO codes gave essentially full spin 
polarization but was not
able to open a gap\cite{Pickett_bnoo} with a reasonable value of U.  
The complication is that the occupied orbital is
a OsO$_6$ cluster orbital with half of the charge on Os and the other half spread over the neighboring
O ions. U should be a value appropriate to this cluster orbital and should be applied to that orbital,
however the codes applied U only to the Os $5d$ orbitals. Xiang and Whangbo\cite{Whangbo2007} neglected
the Hund's rule $J_H$ in the DFT+U method, and did obtain a gap. However, neglecting $J_H$ omits both
the Hund's rule exchange energy and the anisotropy (orbital dependence) of the Hubbard interaction
$U_{mm'}$, whereas one of our intentions is to include all orbital dependencies to understand the
anisotropy on the Os site.  

In this paper we first establish how to model this system faithfully including all anisotropy,
 then address the interplay of SOC
coupling with correlation effects and crystal field splitting.  The density functional extension to
include some fraction of Hartree-Fock exchange -- the hybrid functional -- open the gap, but only when SOC is included.
The inclusion of nonlocal exchange and SOC  provides an understanding of the Mott insulating
ground state and a [110] easy axis, both in agreement with experimental data.\cite{fisherbnoo}
We conclude that BNOO provides an example of a $J_{\rm eff} = \frac{3}{2}$, quarter-filled
Mott insulator. Some comparison is made to isovalent BLOO, which had 6\% smaller volume and aligns
antiferromagnetically rather than ferromagnetically. 

\section{Previous theoretical work}
Two density functional theory based studies, mentioned briefly above, 
have been reported for BNOO. One was performed within a fully anisotropic implementation of the
DFT+U method\cite{Pickett_bnoo} while the other was 
DFT+U\cite{Whangbo2007} (GGA+U), but neglecting anisotropy
of the interaction.  
An overriding feature of this system is a strong hybridization of Os 5$d$ orbitals with O 2$p$ states, with the result 
that the ``$t_{2g}$'' bands have half of their density on the four neighboring oxygen ions
in the plane of the orbital. In keeping
with this strong hybridization, the ``O $2p$'' bands have considerable Os $5d$ charge, such that the $5d$
occupation of the nominally $d^1$ ion is actually 4-5 electrons, still leaving a highly charged ion but
less than half of the formal 7+ designation. 
Lee and Pickett reported\cite{Pickett_bnoo} that fully anisotropic DFT+U 
could not reproduce a Mott
insulating state because U is applied on the Os ion whereas half of the 
occupied local orbital (cluster orbital)
density lies on neighboring oxygen ions.  A model treatment in which U is applied to the cluster orbital
did produce the Mott insulating state.\cite{Pickett_bnoo}
Whangbo's results indicate that part of the complication in this system involves
the anisotropy of the repulsion within the Os ion.

The study reported by Whangbo {\it et. al.}\cite{Whangbo2007} 
addressed three spin directions ([001], [110], and [111]). Within their treatment, 
the [111] spin direction led to the minimum energy, 
indicated a calculated band gap of 0.3 eV for U=0.21 Ry = 2.85 eV (J$_H$=0).  
Using DFT+U+SOC and the same code
but including anisotropy of the interaction $U_{mm'}$, we have
not reproduced this gap, indicating their gap is due to the neglect of Hund's rule coupling and anisotropy
of the interaction. 
Because of the need to include all interactions and all anisotropy, we have
adopted a different approach based on the hybrid exchange-correlation functional, described in the next section. 
This approach seems to be more robust, allowing us to probe the interplay of SOC and strong correlation of BNOO,
and also obtain results of the effect of pressure on the ground state of BNOO.  

Our first challenge was to obtain a Mott insulating state in BNOO when all interactions (correlation and
SOC) are accounted for. With large SOC the result depends on the assigned direction of the
moment.  The hybrid functional approach plus SOC leads directly to a ferromagnetic (FM) Mott insulating ground
state, as observed.\cite{fisherbnoo}
In our studies we observed a strong preference for FM alignment, versus the commonplace antiferromagnetic
(AFM) alignment that often arises on the simple cubic lattice of perovskite oxides. Ba$_2$LiOsO$_6$ however
orders antiferromagnetically, which for nearest neighbor antialignment exchange coupling leads to
frustration of ordering.
All calculations reported here for either compound are for FM orientation.
We study specifically the effects of spin orientation on the electronic structure, and
initiate a study of the pressure dependence of BNOO considering the zero pressure lattice
constant and at 1\%, 2\%, and 5\% reduced lattice constants.

\section{Computational Methods}
The present first-principles DFT-based electronic structure calculations 
were performed using the full-potential augmented plane
wave plus local orbital method as implemented in the WIEN2k code.\cite{wien2k} 
The structural parameters of BNOO with full cubic symmetry of the double
perovskite structure were taken from experimental X-ray
crystallographic data:\cite{crystalbnoo} $a$=8.28~\AA, $x_O$=0.2256.
Non-overlapping atomic sphere radii of 2.50, 2.00, 1.80, and 1.58 a.u. are used for the 
Ba, Na, Os, and O atoms, respectively. The Brillouin zone was sampled with a 
minimum of 400 k points during self-consistency, coarser meshes were sometimes found
to be insufficient.

For the exchange-correlation energy functional for treating strongly correlated insulators,
a variety of approaches in addition to DFT+U exist and have
been tested and compared for a few selected systems.\cite{tran2006} 
As mentioned above, for technical
reasons -- the relevant orbital is an octahedron cluster orbital rather than the
standard localized, atomic-like orbital encountered in $3d$ oxides -- the LDA+U method
is problematic.  We have chosen to apply the onsite exact exchange
for correlated electrons (EECE) functional introduced and evaluated by Novak and
collaborators.\cite{novak2006}, implemented similarly to common use in hybrid
(mixture of Hartree-Fock and local density exchange). This oeeHyb functional 
is an extension of the DFT+U method
to parameter-free form: exact exchange with full anisotropy
is evaluated for correlated orbitals (Os $5d$
orbitals here) without explicit reference to any (screened or unscreened) Hubbard U
repulsion or Hund's exchange interaction J$_H$. The double counting term is evaluated
directly from the density of the occupied correlated orbitals, again without any
input parameters. The exact exchange is calculated within the atomic sphere in atomic-like
fashion (hence ``onsite''). The onsite exact exchange replaces 25\% of the
local density exchange.

This functional is implemented in the Wien2k code,\cite{wien2kinput} and we refer
to it here as oeeHyb.
For the semilocal exchange-correlation functional, the
parametrization of Perdew, Burke, and Ernzerhof\cite{PBE} (generalized gradient
approximation) is used. SOC was included fully in core
states and for valence states was included in a second-variational 
method using scalar relativistic wave functions,\cite{wien2ksoc}  a procedure that
is non-perturbative and quite accurate for $d$ orbitals even with large SOC.

This oeeHyb method has some kinship with hybrid exchange-correlation functionals (see
Tran {\it et al.}\cite{tran2006} for a comparison of several hybrids). Hybrids replace
some fraction $\alpha$, typically 25\%, of local density exchange with Hartree-Fock
exchange, which then is approximated in various ways to reduce the expense to a reasonable
level.  The oeeHyb approach deals with exact exchange only for correlated orbitals,
however, making it appropriate for correlated materials but it will not increase
bandgaps of ionic or covalent semiconductors.


We note that as an alternative to the commonly used DFT+U approach,
the more conventional hybrid exchange-correlation functional as implemented in the 
VASP code\cite{VASP} has been 
applied to the iridate Na$_2$IrO$_3$ by
Kim {\it et al.}\cite{kimHSE} to obtain the magnetic insulating ground state.


\section{Electronic structure and magnetic moments}
\subsection{Electronic structure: dependence on spin direction}

\begin{figure*}
\centering
\includegraphics[width=1.0\linewidth]{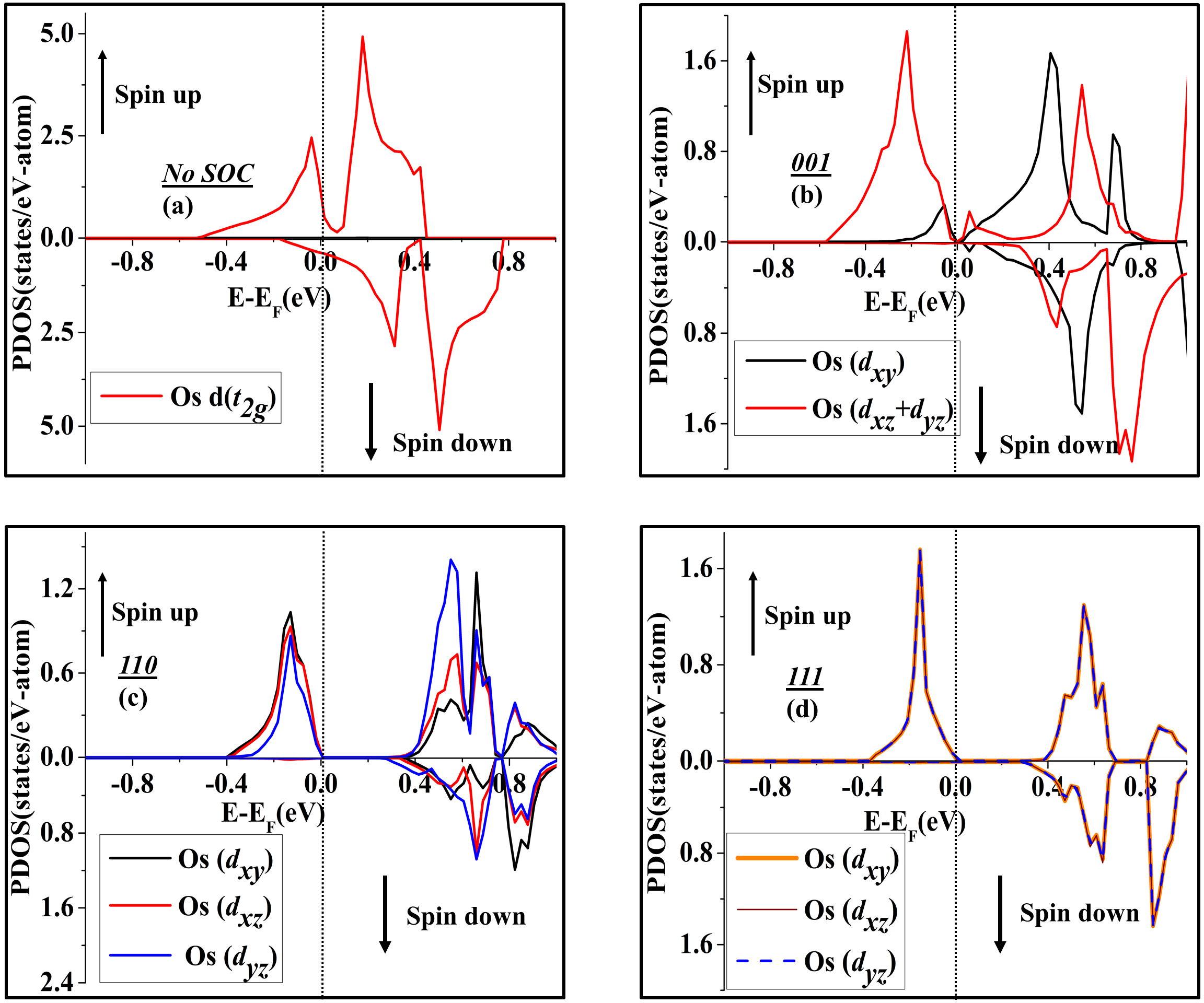}
\caption{Orbital-projected density of states plot for BNOO using the oeeHyb functional (a) without SOC,
and for three different spin orbit coupling directions: (b) [001], (c) [110], and (d) [111]. The 
horizontal axis is in eV relative to the Fermi level (a), or top of the gap (b,c,d). The vertical 
axis is in states/eV-atom/formula unit. The most relevant information is the relative occupations of $t_{2g}$
orbitals in the lower Hubbard band (-0.4 to 0.0 eV).}
\label{dos_fig1}
\end{figure*}

Projected densities of states (PDOSs) are presented in Fig. \ref{dos_fig1} for BNOO at the experimental volume
using oeeHyb, initially without SOC and with  full cubic symmetry. The Fermi
level (taken as the zero of energy) lies in a deep pseudogap, due to small band overlap.
All $t_{2g}$ orbitals participate equally at this level, evident from the observation that the PDOS is
distributed almost equally over the Os $t_{2g}$ orbitals and the $2p$ orbitals of the
six neighboring O ions, a very strong hybridization effect that has been emphasized
before.\cite{Pickett_bnoo}

The conventional picture of a Mott insulating state in a $t^1_{2g}$ shell is that a single orbital,
say $d_{xy}$, is occupied, and the crystal symmetry is broken (and must be broken in the calculation)
to sustain, indeed to allow, occupation of a single orbital.  However, due to the large SOC which is
expected to produce a substantial orbital moment which requires occupation of a complex 
({\it viz.} $d_{xz} \pm i d_{yz}$) orbital,
we have foregone this intermediate step of obtaining orbital-ordering broken symmetry, 
which would typically be the final result for a $3d$ ion
with negligible SOC. 

Adding SOC to oeeHyb, the  result mentioned above, which necessarily lowers the symmetry, 
leads to a SOC-driven Mott insulating state with a 0.2-0.3 eV gap, 
depending on the direction assumed for the magnetization.
It is for this reason that we use the oeeHyb functional to model BNOO, as we have 
previously reported that the LDA+U approach was unable to open a gap.\cite{Pickett_bnoo} 
The GGA+U method is more promising, but it requires an 
unphysically large value of $U$ = 6 eV to open a gap for [001] orientation. 
One can obtain a band gap along for [110] and [111] orientations in GGA+U with U values of 
3.2 eV and 2.0 eV, respectively.\cite{kwlee}

Our calculated band gaps, provided in Table \ref{Table1}, depend on direction of the
magnetization, which experimentally can be manipulated with an applied field.
It should be kept in mind that symmetry is lowered by including SOC, and the
resulting symmetry depends on the direction of magnetization. 
For spin along [001] the band gap is lower by 70 meV than for [110] and [111].

 \begin{table}
 \begin{tabular}{c c c c c c c c c}
 \hline Method &  $\mu_{s} $ &  $\mu_{l}$ & $\mu_{tot} $ &  Band Gap  \\
 \hline
 \hline Lattice parameter 8.28 \AA  \\
 \hline
 oeeHyb & 0.59 & N/A  & N/A   &  none \\
 oeeHyb+SOC (001) & 0.52 & -0.41  & 0.11 & 0.02 \\
 oeeHyb +SOC (110) & 0.52  &-0.44  & 0.08  & 0.28  \\
 oeeHyb +SOC (111)  & 0.52 & -0.45 & 0.07  & 0.30 \\
 \hline Lattice parameter 8.20 \AA  \\
 \hline
 oeeHyb & 0.53  & N/A  & N/A   &  none \\
 oeeHyb+SOC (001) & 0.49  & -0.37  & 0.11  & none \\
 oeeHyb +SOC (110) & 0.48 & -0.41  & 0.08  & 0.21 \\
 oeeHyb +SOC (111)  & 0.48  & -0.42 & 0.06 & 0.26  \\
 \hline lattice parameter 8.10 \AA \\
 \hline
 oeeHyb & 0.51  & N/A  & N/A   &  none  \\
 oeeHyb+SOC (001) & 0.48  & -0.37  & 0.12 & none\\
 oeeHyb +SOC (110) & 0.48 &  -0.27 & 0.21 & 0.12 \\
 oeeHyb +SOC (111)  & 0.48 & -0.35 &  0.13 & 0.15 \\
 \hline Lattice parameter 7.86 \AA  \\
 \hline
 oeeHyb& 0.48 & N/A  & N/A  & none \\
 oeeHyb +SOC (001) & 0.46  & -0.34  & 0.12 & none \\
 oeeHyb +SOC (110) & 0.47  & -0.38   & 0.09& none \\
 oeeHyb +SOC (111)  & 0.47 & -0.39 & 0.09 & none \\
 \hline
 \end{tabular}
 \caption{Calculated spin, orbital, and total ($\mu_{tot}$=$\mu_s$+$\mu_{\ell}$) moments of Os,
and band gap of Ba$_2$NaOsO$_6$, for four values of lattice parameter and for the three
high symmetry directions of the magnetization. Note: the total moment per f.u. will
include an $\sim$0.5$\mu_B$ spin moment not included within the Os sphere,
thus primarily on the oxygen ions.}
\label{Table1}
 \end{table}

In the $d^1$ Mott insulating state a single (Wannier) orbital is occupied. While SOC mixes
spins, the splitting the lower (majority) and upper (minority) Hubbard bands is sufficiently
large that the system remains essentially fully spin-polarized. The moment on Os is
decreased by 0.07 $\mu_B$ (see Table 1) by the rebonding induced by SOC. The orbital moment
is  around -0.4 $\mu_B$ for all directions of the moment, with the difference from unity being
primarily due to half of the Wannier function lying on the O ions.

The PDOSs displayed in Fig.~\ref{dos_fig1}
indicate the character of the occupied orbital, and this figure clearly illustrates the large effect of
SOC on the $t_{2g}$ spectrum. For [001] spin direction, the dominance of
$d_{xz}, d_{yz}$ orbitals in the PDOS, along with the orbital moment reported in 
Table \ref{Table1}, indicates occupation of the $d_{xz} -id_{yz}$ = $d_{m=-1}$ orbital,
with moment reduced due to the strong hybridization (hence partial
quenching) with O $2p$ orbitals. For [110], again $d_{xz}, d_{yz}$ are equally occupied,
however $d_{xy}$ contributes somewhat more. For [111], all $t_{2g}$ orbitals contribute
equally, reflecting no evident symmetry breaking beyond that of choosing a specific
[111] axis for the direction of magnetization. The occupied bandwidth in each case is
0.4 eV, though the differing shape of the band for the different directions of spin
(discussed below) reflects the lowering of symmetry of the bands by SOC. 

The relevant bands near the gap along selected symmetry directions
are shown in Fig.~\ref{fig:BNOO_BLOO_810} for each of the three
directions of spin. These plots are shown for the 2\% reduced lattice constant 
$a$=8.10~\AA, for later comparison with BLOO.  At the experimental volume, the results are extremely similar except
that, for the [001] direction, the bands along W-K do not cross, leaving a small gap.
The bands are quite flat through much of the zone (for example, along W-L-$\Gamma$ directions), 
with important dispersion around the X point, and also
dispersion along the W-K directions that is spin-direction dependent. 

\subsection{Magnetocrystalline anisotropy}
The energy difference 
between [110] and [111] spin directions is small, and the band gaps for these two directions are 
indistinguishable. Fig.~\ref{dos_fig1}(b,c) illustrates the $t_{2g}$ symmetry breaking due to
SOC: for [001] $d_{xz}$ and $d_{yz}$ maintain equal occupation and dominate, each contributing 40\% 
and obviously providing the orbital moment. There is also 20\% $d_{xy}$, 
but without $d_{x^2-y2}$ it cannot
contribute to the orbital moment.
For the easy axis [110] spin direction $d_{xy}$ is also distinct from the other 
two, but interpreting the
orbital moment in this global basis requires rotations with knowledge of phases. 
For the [111] spin direction 
all three orbitals contribute equally in
the lower Hubbard band by symmetry.  

From the calculated energies we obtain [110] as the easy axis, as determined 
experimentally.\cite{fisherbnoo} The [111] direction is very close in energy 
however ($\sim$1 meV, close to our precision), while the [001] direction is
13 meV higher.
This agreement with experiment of the easy axis, following our success in 
reproducing the FM Mott insulating ground state, 
is an important validation
of using the hybrid functional to model BNOO. The atom-
and orbital-projected densities of
states plot of Fig.~\ref{dos_fig1} does not indicate directly the strong 
participation of O $2p$ orbitals in the
nominal Os $5d$ bands, but this aspect has been emphasized in the earlier 
studies.\cite{Pickett_bnoo,Whangbo2007}

\begin{figure}
\centering
\includegraphics[width=0.9\linewidth]{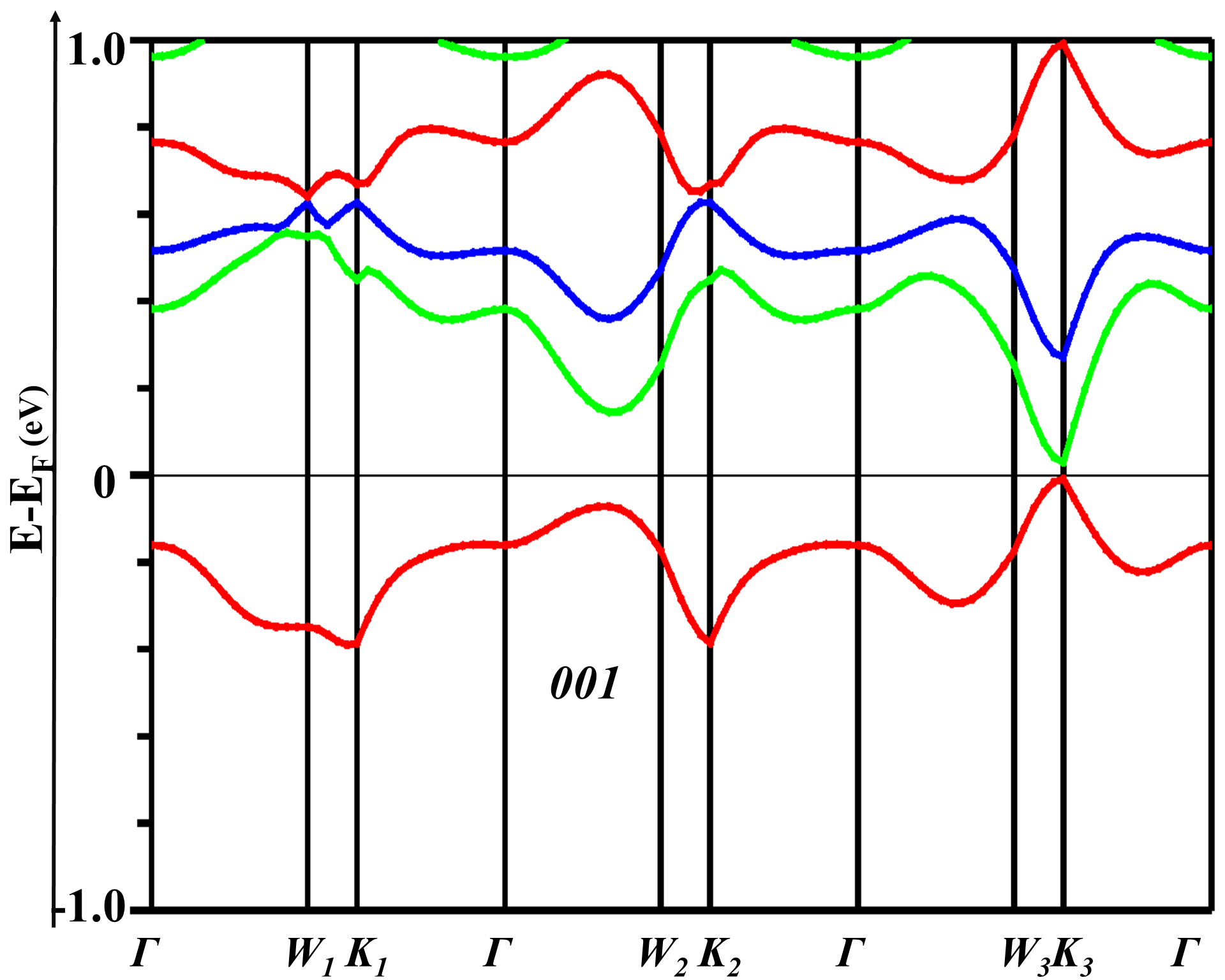}
\caption{Band plot along three different W-K directions for Ba$_2$NaOsO$_6$ with
[001] magnetization axis ($a$= 8.28~\AA), illustrating the large effect of symmetry
lowering by the large spin-orbit interaction strength. The chosen points, in units of
$\frac{\pi}{a}$, are $\Gamma$=(0, 0, 0),
W$_1$=($\frac{1}{4}$, $\frac{1}{4}$, $\frac{1}{2}$),
W$_2$=(-$\frac{1}{2}$, $\frac{1}{4}$, -$\frac{1}{4}$),
W$_3$=($\frac{1}{2}$, $\frac{1}{4}$, -$\frac{1}{4}$),
K$_1$=($\frac{3}{8}$, 0, $\frac{3}{8}$),
K$_2$=($\frac{3}{8}$, $\frac{3}{8}$, 0),
K$_3$=($\frac{3}{8}$, 0, -$\frac{3}{8}$).}
\label{001wk}
\end{figure}

\begin{figure*}[tb!]
\centering
\includegraphics[width=0.9\linewidth]{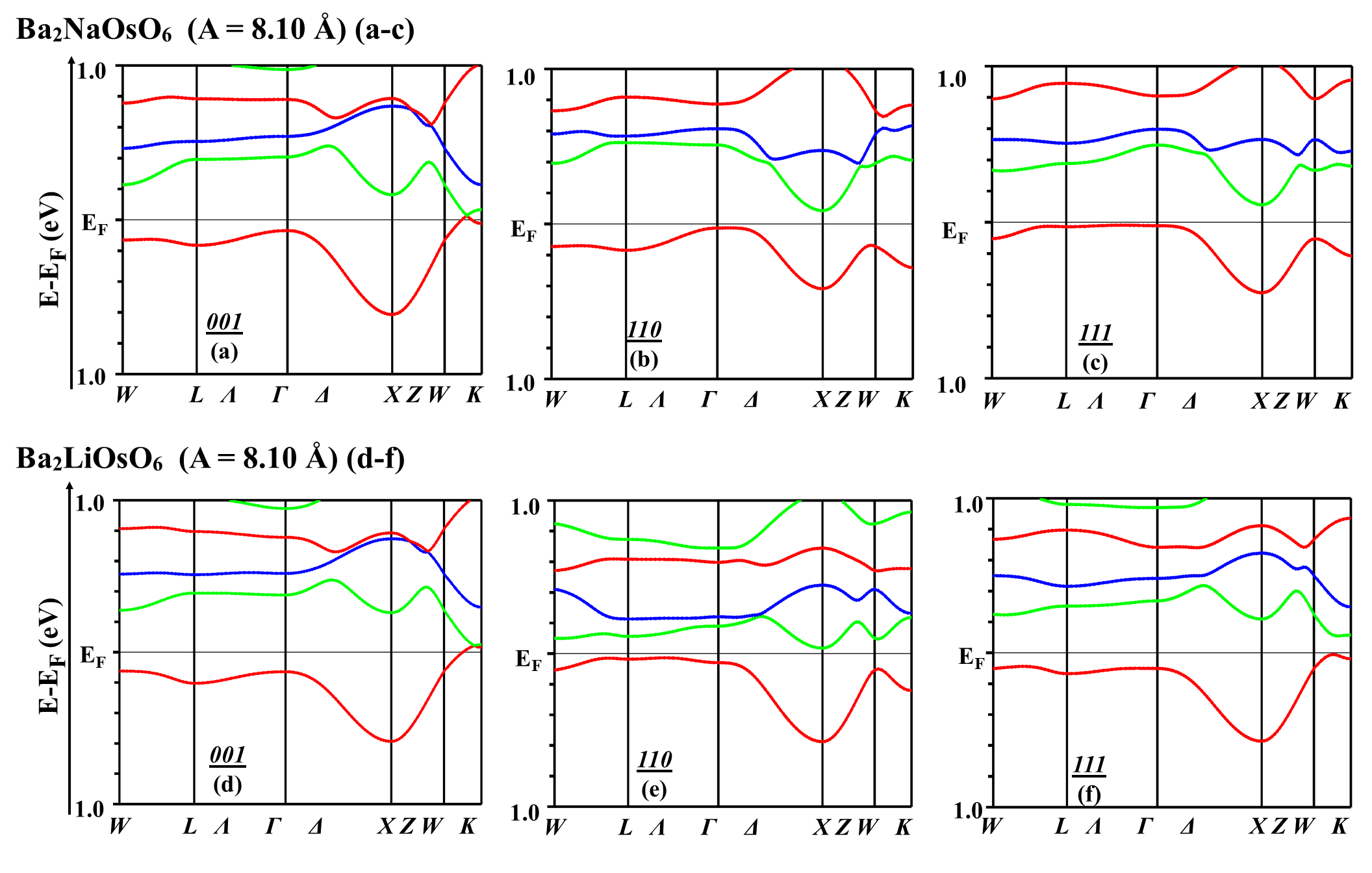}
\caption{Band plots for Ba$_2$NaOsO$_6$ (a-c) and Ba$_2$LiOsO$_6$ (d-f) for the
three directions of magnetization axis (indicated on the plots)
and with a lattice parameter of 8.10 ~\AA~
(experimental lattice parameter of Ba$_2$LiOsO$_6$). The differences are
quite small for [001] but are important along W-X-K for the other two spin directions.}
\label{fig:BNOO_BLOO_810}
\end{figure*}

Including SOC lowers symmetry substantially so there are, for example, several
different ``X-W-K" directions along the edge of the zone.  
Fig.~\ref{fig:BNOO_BLOO_810} illustrates that there is actually a substantial 
change of dispersion
along the zone boundary X-W-K directions, which disperse very differently for [001] compared 
to [111] and [110] spin directions.
Otherwise the splitting of bands remains very similar, roughly 0.4 eV throughout the zone. 
In Fig.~\ref{001wk} the band plot for BNOO is shown along three inequivalent $\Gamma$-W-K directions
for [001] magnetization direction.  Along the W$_3$-K$_3$ direction the band gap valence nearly vanishes
(15 meV gap). With a decrease in lattice parameter, eventually the gap collapses, yielding an
insulator-to-metal transition at a critical pressure, which is spin-direction dependent.

\begin{figure*}
\centering
\includegraphics[width=0.9\linewidth]{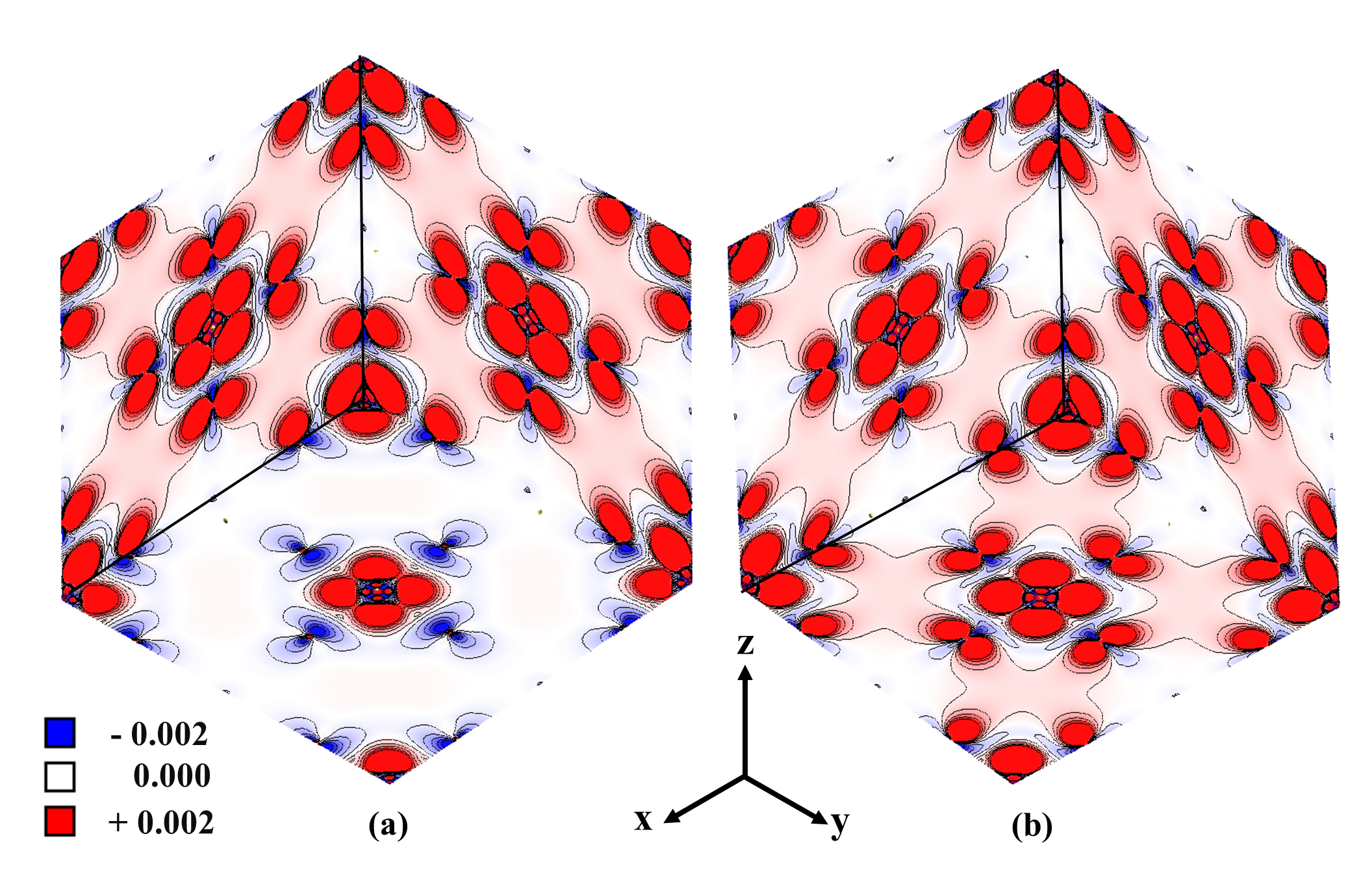}
\caption{Spin density contour plots in three planes, for Ba$_2$NaOsO$_6$ 
using the hybrid functional. Panel (a) 
corresponds to the [001] spin direction, panel (b) to the [111] spin direction.
Os ions lie at the centers of the faces and at the corners, neighbored by four O ions
in these planes. The blue contours in panel (a) reflect the negative spin polarization on
the $2p_{\sigma}$ orbitals in the equatorial plane. See also Fig.~\ref{fig:Slide1}.
The figure was produced with the Xcrysden software.\cite{xcrysden}}
\label{fig:sdens_001_111}
\end{figure*}

\subsection{Pressure dependence}
To access the effect of pressure we reduced the experimental lattice parameter of BNOO by 0.08~\AA, 
0.18~\AA, and 0.42~\AA (1\%, 2\%, and 5\% respectively) 
and repeated hybrid DFT calculations for the three magnetization directions. 
Table \ref{Table1} provides the changes of the
calculated band gap. For [001] orientation, band overlap and thus a 
insulator-metal transition (IMT) occurs. However, 
for [111] and [110] orientation the gap remains, having decreased by 30\%. 
The relative energy difference 
is increased by the emergence of the metallic state, becoming several times higher 
than for the insulating states. From the
bands plotted in a few symmetry directions in Fig. 3, 
collapse of the gap {\it per se} is not the reason 
for the IMT, rather it is the dispersion related to the direction of magnetization. 

The dependence of the band dispersion on magnetization direction that is shown in
Fig.~\ref{fig:BNOO_BLOO_810} reveals a new kind of insulator-metal
transition, one in which the tuning parameter is the magnetization direction. This
direction can be manipulated by a sufficiently large applied field. Taken at
face value, our modeling predicts that this type of transition will occur in BNOO, with
an onset with pressure (decreasing volume) slightly below the experimental volume. The
necessary pressure can be estimated using our calculated bulk modulus of B = 50 GPa.
Interpolating between our gap values at 8.28~\AA~and 8.10~\AA~(0.02 eV and -0.10 eV,
respectively) give gap closing at $|\Delta a|/a \sim$ 4$\times$10$^{-3}$, 
thus a gap closing and insulator metal transition in
the vicinity of 5 kbar. Our characterization above 
of ``at face value'' refers to the well known
issue that band gap values are uncertain at the level of several tenths of eV even when
correlation corrections are made, and that the relation  of oeeHyb band gaps to 
experimental values in correlated oxides is not established.

\subsection{Similarities and differences: BNOO vs. BLOO}
BLOO, the isostructural ($a$=8.10~\AA, $x_O$=0.2330)\cite{crystalbnoo} and isovalent 
sister compound of BNOO, has an AFM ground state versus FM in BNOO. 
The primary difference seems to be the Os-O distance, 1.81~\AA versus 1.87~\AA in BNOO.
In the model of Chen \cite{balentsprb1} {\it et al.}that includes both intersite spin exchange J' and intersite
Coulomb repulsion V, this structural difference suggests a larger V for BLOO, Given the 
similarities of the two band structures (thus, hopping amplitudes) and expected similarity
of correlation strengths, J' may be the same for both compounds.   

With strong SOC, the mechanism of magnetic ordering involves the tensorial coupling of spin+orbital
moments,\cite{STPi} which lies beyond the scope of this paper. For purposes of comparison,
we have performed all calculations for BLOO keeping ferromagnetic orientation to enable close comparison 
with BNOO.  In Fig. \ref{fig:BNOO_BLOO_810} it is shown that, at a given volume, 
the band structures are similar over most of the zone. Because
the gap is small, there are differences on the fine scale that are significant. 
At the experimental volume of BLOO,
there is small band overlap along the plotted W-K direction for [001] magnetization direction for BNOO versus
band touching (zero gap) for BLOO. Generally there are only visible differences in the vicinity of the W-K
direction, which holds for other volumes that were studied.  

It was mentioned above that the calculated easy axis of BNOO is [110], with [111] being 
close.  This similarity of [110] and [111] energies persist for smaller lattice
constants. The higher energy of [001] becomes even larger when the electronic
structure becomes metallic, {\it i.e.} beyond the insulator-metal transition. 
For BLOO the easy axis (assuming FM alignment) changes with volume: 
at the experimental volume [111] is 
the easy axis but at the BNOO lattice parameter ({\it i.e.} expanded by 0.18 ~\AA) [001] 
becomes preferred. 
We note that the charge on Os is the same for both BNOO and BLOO for all volumes studied.

It can be seen, by comparing Table \ref{Table2} with Table \ref{Table1},
that the spin, orbital, and total moments on Os are affected somewhat by the Li/Na difference,
consistent with the different behaviors seen experimentally..
The differences can be appreciable and are both volume and spin-direction dependent. 
For example,
the moment compensation is greatest for [111] polarization for both compounds 
($\mu_{tot}$= 0.07-0.08$\mu_B$), but the greatest compensation occurs at
different volumes (each at its own volume).
There are differences in the band structures of BNOO and BLOO on the 0.1 eV scale, 
with BLOO generally having a somewhat
smaller band gap. From Fig. \ref{fig:BNOO_BLOO_810} it is evident that the band
structures differ  strongly along several of the W-K directions.  

\begin{figure*}
\centering
\includegraphics[width=0.9\linewidth]{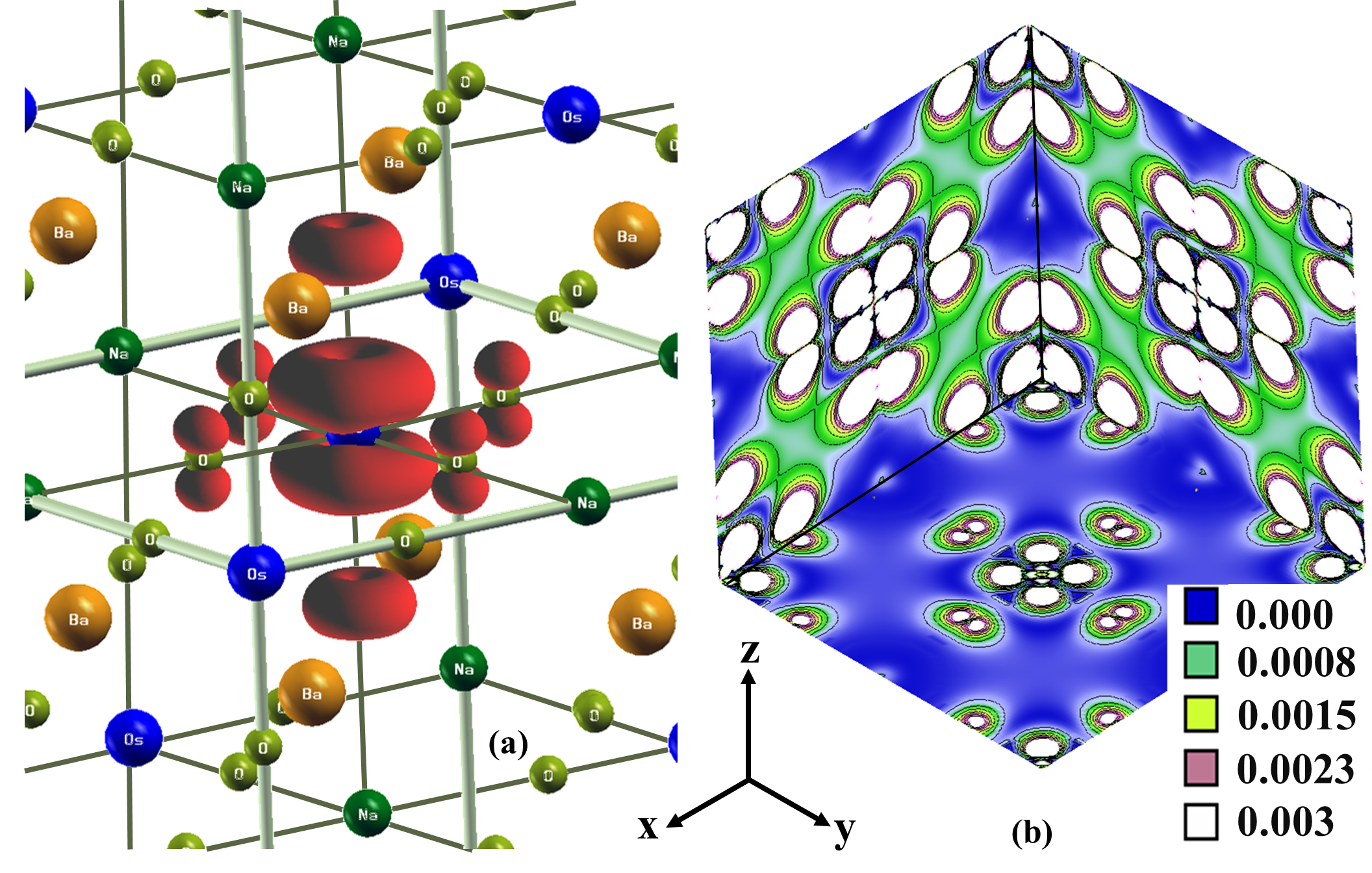}
\caption{Left panel: isocontour plot of the spin density in the limited range
-04 eV to 0.0 eV -- the occupied lower Hubbard band -- 
for Ba$_2$NaOsO$_6$ at its experimental volume and [001] spin direction
and using the hybrid DFT 
functional.  This spin density is non-negative, in line with lack of any occupied
minority spin, see Fig.~\ref{dos_fig1}. Right panel: isocontour plot of the
same density, showing more detail. For example, the small spin density in the
$x-y$ plane does not show up on the isocontour plot. The negative spin density on the
equatorial O ions that is evident in Fig.~\ref{fig:sdens_001_111} therefore
arises due to the polarization of the O $2p$ bands (with no net moment).}
\label{fig:Slide1}
\end{figure*}

\section{Dependence of spin density on magnetization direction}
The PDOS plot of BNOO shows how degeneracy of Os $t_{2g}$ orbitals 
(Fig. \ref{dos_fig1}(a)) is broken when SOC is included. As already mentioned,
Fig. \ref{dos_fig1}(b) for [001] orientation shows that $d_{xz}$ and $d_{yz}$ predominates 
the occupied band, with their complex
combination ($m$=~-1) accounting for the orbital moment. For illustration, 
Fig.~\ref{fig:sdens_001_111}(a,b) provides plots of the
spin density contours in the three basal planes for [001] and [111] magnetization directions, to
illustrate the differences.
The spin density contours for [001] spin direction in the $y-z$ and 
$x-z$ planes are clearly very different from those
in the $x-y$ plane, indicating its {\it strong anisotropy}. 
For the [111] orientation shown in Fig. \ref{fig:sdens_001_111}(b), 
the spin density contours are equivalent in
the three planes, reflecting the expected threefold symmetry. Some other symmetries
will be absent, however. 

Fig.\ref{fig:Slide1} (left panel) provides a 3D isosurface plot as well as spin density 
contours (right panel) for the lower Hubbard band 
(occupied region at -0.4 eV to 0.0 eV)  of BNOO at its ambient pressure volume,
 for [001] magnetization direction. 
The left panel reveals the $d_{xz} - i d_{yz}$ spin density driven by SOC,
which appears as large donuts around the Os site.
The figure also shows the $p_x - i p_y$ spin density on the apical oxygen ions,
which is driven by $p-d$ hybridization rather than direct SOC on oxygen
(which is very weak for Z=8).
This plot reinforces the earlier inferences about strong cubic symmetry breaking
and additionally it shows that the contribution of $d_{xz}$, $d_{yz}$ orbitals 
dominates the $d_{xy}$ contribution. 

The corresponding contour plot in Fig. \ref{fig:Slide1} (right panel)
provides more detail about the Os $d_{xz}$, $d_{yz}$ orbitals hybridizing with $p_{z}$ orbital 
of apical O, which thereby acquires an orbital moment. The in-plane O $p_{z}$ orbital
(a p$_{\pi}$ orbital) has substantial participation in 
bonding, but the $p_z$ orbital contributes no orbital moment.

\begin{figure}
\centering
\includegraphics[width=0.9\linewidth]{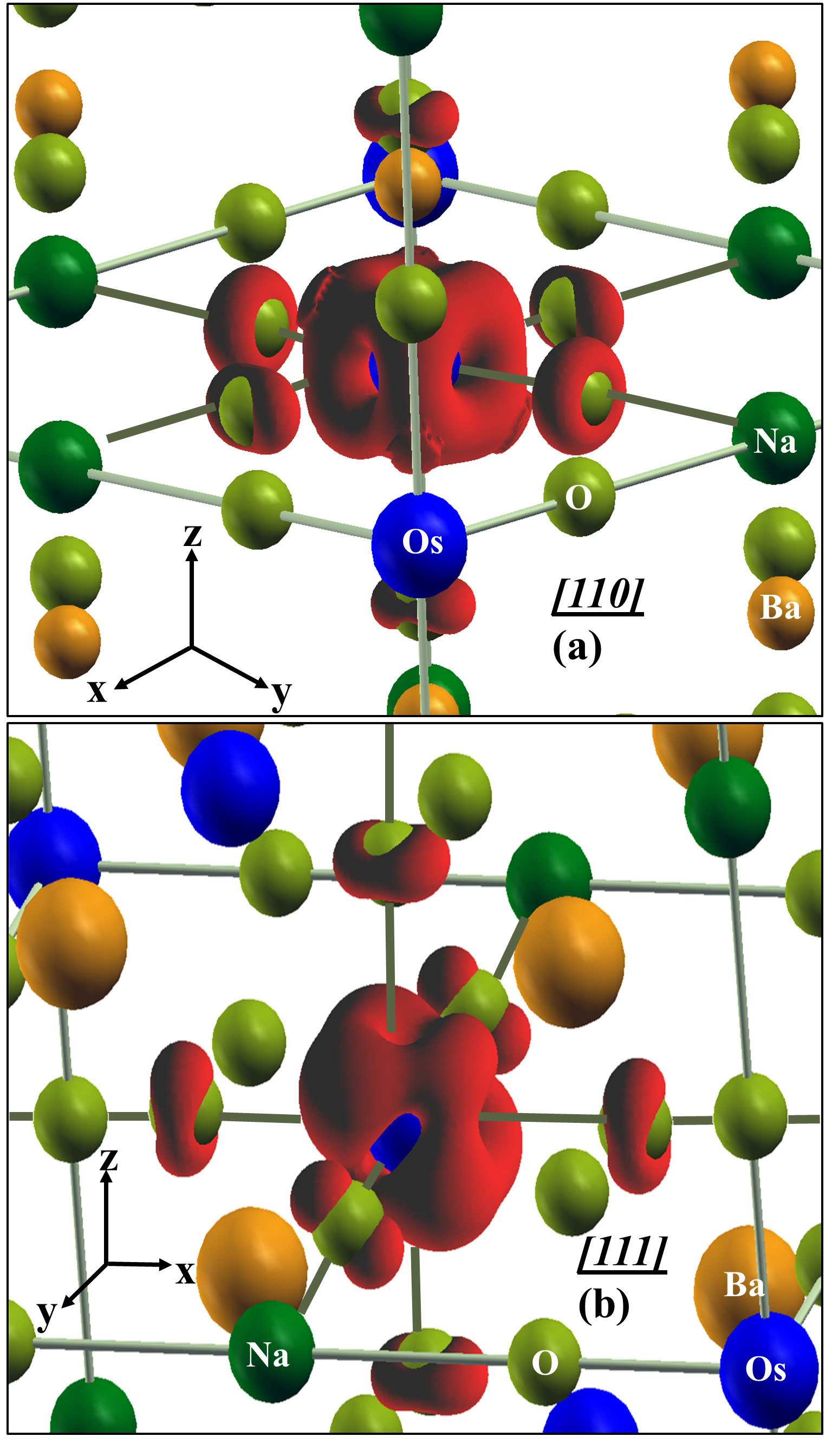}
\caption{Isosurface plot of the spin density of BNOO as in Fig.~\ref{fig:Slide1},
but for [110] (top panel) and [111] (bottom panel) spin orientation. Note the
near-cubic symmetry of the [110] density on both the Os ion and the neighboring
O ions. The [111] density shows three-fold symmetry around the chosen [111]
axis for an $m_{\ell} = \pm 1$ orbital for Eq.~\ref{eqn1}. 
}
\label{fig6}
\end{figure}

Finally, in Fig.~\ref{fig6} the spin density (which contributes the anisotropic charge density)
is provided for both [110] and [111] orientations. For [111], the orbital is the 
$m_{\ell}=-1$ member of the symmetrized combinations of $t_{2g}$ orbitals given by
\begin{eqnarray}
\psi_m = \frac{1}{\sqrt{3}}(\zeta_m^0 d_{xy} + \zeta_m^1 d_{yz} + \zeta_m^2 d_{zx}),
\label{eqn1}
\end{eqnarray}
where $\zeta_m = e^{2\pi mi/3}$ is the associated phase factor for threefold rotations.
The spin density is anisotropic but not as strongly so as for [001] orientation.

Less intuitive and more surprising is the spin density for [110], also displayed in
Fig.~6. By sight, it is very nearly cubic; the complex linear combination of $t_{2g}$
orbitals necessary to provide the orbital moment is not at all evident in the spin
density, which is a cube with rounded corners and dimpled faces. The spin density on
the O ions in unremarkable and also nearly symmetric.   
This near-cubic symmetry provides an explanation for the lack of an observable
Jahn-Teller distortion in this $d^1$ system.


\subsection{Comments on magnetic moments} 
The orbital moment of Os is around -0.4$\mu_B$, with spin moment on Os about 0.5$\mu_B$, for all volumes 
and spin directions we have studied. Thus the net moment is very small, only $0.1\mu_B$, on Os, 
and {\it most of the net (and observed) moment 
of} 0.6$\mu_B$ {\it resides on the neighboring O ions}. For the [001] spin direction, 
the apical O acquires a spin moment three times larger than those in the equatorial 
plane, while for [110] the equatorial O ions have 1.5 larger spin.The spin density 
isocontours for spin along [001], shown in Fig \ref{fig:Slide1}, reveals its strong 
non-cubic symmetry. 

Another unexpected feature arises from the calculation.
For spin along [001], the apical O ion acquires an orbital moment approaching 0.1 $\mu_B$, 
whereas the in-plane O ions show negligible orbital moment.  
We are not aware of a significant
orbital moment being detected, or predicted, on any oxygen ion. 
XMCD is the conventional measurement for obtaining the
orbital moment, though tying it to the O ion will be challenging because 
there is substantial Os $5d$ character mixed
with O $2p$, and vice versa.  Detection of an orbital moment on O will help to substantiate 
the modeling of this unusual
large SOC, $J_{\rm eff}$ = $\frac{3}{2}$ FM Mott insulator by the hybrid exchange functional,
as well as generalize expectations of where orbital magnetism may emerge.
 
 \begin{table}
 \begin{tabular}{c c c c c c c c c}
 \hline Method &  $\mu_{s} $ &  $\mu_{l}$ & $\mu_{tot} $ &  Band Gap  \\
 \hline
 \hline Lattice parameter 8.28 \AA  \\
 \hline
 oeeHyb    & 0.62 & N/A &  N/A  & none \\
 oeeHyb+SOC (001) & 0.53 & -0.40 & 0.13   & 0.02 \\
 oeeHyb+SOC (110) & 0.56  &-0.26 & 0.31& 0.13 \\
 oeeHyb +SOC (111)  & 0.53 & -0.32  & 0.22  & 0.09 \\
 \hline Lattice parameter 8.10 \AA  \\
 \hline
 oeeHyb & 0.60 & N/A  & N/A     & none \\
 oeeHyb+SOC (001) & 0.53  & -0.39 & 0.13  & none \\
 oeeHyb  +SOC (110) & 0.52 & -0.41  & 0.11  & 0.05 \\
 oeeHyb +SOC (111)  & 0.52 & -0.44  & 0.08  & 0.06  \\ 
 \hline 
 \end{tabular}
 \caption{Analogous to Table \ref{Table1}, but for Ba$_2$LiOsO$_6$ and for only two lattice constants.
Note: the total moment per f.u. will
include an $\sim$0.55$\mu_B$ spin moment not included within the Os sphere,
thus primarily on the oxygen ions.}
 \label{Table2}
 \end{table}
    
 \section{Summary}
The hybrid density functional has been used for the modeling of Ba$_2$NaOsO$_6$ with considerable success. 
Unlike our attempts with GGA+U+SOC, when SOC is included, and only then,
it provides a robust FM Mott insulating ground state for the moment
aligned along any of the three symmetry directions. Thus BNOO (and BLOO) is a $J_{\rm eff}$=$\frac{3}{2}$
Mott insulator at quarter filling, driven by the combination of strong exchange-correlation and large
SOC. Though the spin magnetization remains completely spin-up when SOC is included, the large
changes in the composition of the occupied orbital are driven by SOC. 
On the scale of 0.1-0.2 eV which is important considering the small gap, the band structure is
strongly dependent of the orientation of the magnetization.

In addition, our approach predicts the [110] direction as the
easy axis, as observed, though [111] is close in energy. 
The character  of the net moment is unexpected. On the one hand,
the spin and orbital moments on Os are 0.5 and -0.4$\mu_B$, respectively, leaving a net moment on
Os of 0.1$\mu_B$, similar to earlier indications.\cite{Pickett_bnoo,Whangbo2007} 
However this value leaves most of the
total moment of 0.6$\mu_B$ arising from the oxygen ions in the OsO$_6$ cluster. 
The observed ordered moment is ~$\sim$0.2$\mu_B$ thus indicates somewhat larger compensation, or
small spin contribution, than our results provide.

Another unusual feature
found here is that for spin along [001], there is a 0.1$\mu_B$ orbital moment on the 
apical oxygen ion. This surprisingly large value cannot arise from the very small SOC on
oxygen; rather it is the hybridization of the Os $d_{xz} - i d_{yz}$ combination with
the $p_x -i p_y$ orbital that transfers orbital angular momentum to O. This effect does
not operate for other spin orientations.

\section{Acknowledgments}
We acknowledge many useful conversations with K.-W. Lee, discussions with
I. R. Fisher, R. T. Scalettar, and N. J. Curro, and comments on the
manuscript from R. T. Scalettar.
Our research used resources of the National Energy Research Scientific Computing 
Center (NERSC), a DOE Office of Science User Facility supported by the Office of 
Science of the U.S. Department of Energy under Contract No. DE-AC02-05CH11231.
This research was supported by DOE Stockpile Stewardship Academic Alliance
Program under Grant DE-FG03-03NA00071.


\begin{thebibliography}{27}
\bibitem{wien2kinput}In the Wien2k {\rm case.inorb} file the flag {\rm HYBR} is used
for Os $5d$ orbitals, with a
fraction of 0.25.
\end{thebibliography}
\end{document}